\documentclass[aps,pra,twocolumn,showpacs,preprintnumbers,amsmath,amssymb,footinbib]{revtex4}
\usepackage{mathtools}
\usepackage{lipsum}
\usepackage{mathtools,amssymb,lipsum, nccmath}
\usepackage{physics}

\DeclareMathAlphabet{\mathcalligra}{T1}{calligra}{m}{n}
\DeclareMathAlphabet{\mathpzc}{OT1}{pzc}{m}{it}

\usepackage{graphicx,epsfig}
\usepackage{bm}
\usepackage{dcolumn}
\usepackage{xcolor}
\usepackage[breaklinks=true,colorlinks,citecolor=blue,linkcolor=blue,urlcolor=blue]{hyperref}

\def\nat#1#2#3{Nature {\bf #1}, #2 (#3)}

\def\sc#1#2#3{Science {\bf #1}, #2 (#3)}

\def\rmp#1#2#3{Rev. Mod. Phys. {\bf #1}, #2 (#3)}
\def\prl#1#2#3{Phys. Rev. Lett. {\bf #1}, #2 (#3)}
\def\pra#1#2#3{Phys. Rev. A {\bf #1}, #2 (#3)}
\def\prb#1#2#3{Phys. Rev. B {\bf #1}, #2 (#3)}

\def\epjd#1#2#3{Eur. Phys. J. D {\bf #1}, #2 (#3)}

\def\jpb#1#2#3{J. Phys. B: At. Mol. Opt. Phys. {\bf #1}, #2 (#3)}
\def\jpcm#1#2#3{J. Phys.: Condens. Matter {\bf #1}, #2 (#3)}

\def\pla#1#2#3{Phys. Lett. A {\bf #1}, #2 (#3)}

\def\ejp#1#2#3{Eur. J. Phys. {\bf #1}, #2 (#3)}

\def\noi{\noindent}
\def\bc{\begin{center}}
	\def\ec{\end{center}}
\newcommand{\bea}{\begin{equation}}
\newcommand{\eea}{\end{equation}\noi}
\newcommand{\ber}{\begin{eqnarray}}
\newcommand{\eer}{\end{eqnarray}\noi}
\begin{document}
\title{Scaling theory for the collapse of a trapped Bose gas in a synthetic magnetic field: a critical study at the condensation point}
	
\author{Bikram Keshari Behera}
\author{Shyamal Biswas}\email{sbsp@uohyd.ac.in}
\affiliation{School of Physics, University of Hyderabad, C.R. Rao Road, Gachibowli, Hyderabad-500046, India}
	
\date{\today}
	
\begin{abstract}
We have analytically explored both the zero temperature and the finite temperature scaling theory for the collapse of an attractively interacting 3-D harmonically trapped Bose gas in a synthetic magnetic field. We have considered short ranged (contact) attractive inter-particle interactions and Hartree-Fock approximation for the same. We have separately studied the collapse of both the condensate and the thermal cloud below and above the condensation point, respectively. We have obtained an anisotropy, artificial magnetic field, and temperature dependent critical number of particles for the collapse of the condensate. We have found a dramatic change in the critical exponent (from $\alpha=1$ to $0$) of the specific heat ($C_v\propto|T-T_c|^{\alpha}$) when the thermal cloud is about to collapse with the critical number of particles ($N=N_c$) just below and above the condensation point. All the results obtained by us below and around the condensation point are experimentally testable within the present-day experimental set-up for the ultracold systems in the magneto-optical traps.
\end{abstract}
	

	
\maketitle	

\section{Introduction}
The observation of the Bose-Einstein condensation \cite{Anderson,Davis,Bradley} as well as the collapse of an attractively interacting 3-D harmonically trapped Bose-Einstein condensate (BEC) \cite{Roberts} has drawn extensive attention of both the experimentalists and the theoreticians on the ultra-cold systems in magneto-optical traps over the last two decades \cite{Dalfovo,Pitaevskii,Stock,Bloch,Giorgini,Fetter,Dalibard,Goldman,Proukakis}. The collapse of the BEC has a similarity with the gravitational collapse of a white-dwarf star \cite{Gerton}. While essentially the outward Pauli pressure of the electrons (fermions) and the inward pressure due to the gravitation pull of He$^{++}$ ions compete for the stability of a white-dwarf star \cite{Chandrasekhar}, the outward quantum pressure due to the zero-point motion of the condensate and the inward pressure due to the attractive interactions of the bosons compete for the stability of the condensate \cite{Gerton,Biswas7}. The attractive interactions dominate beyond a critical number of particles and cause the collapse of either of the systems \cite{Gerton}. The Bose-Einstein condensate essentially becomes metastable due to attractive interactions and undergoes a macroscopic quantum tunnelling resulting in the collapse for the number of particles close to a critical number $N_c$ \cite{Ueda}.

The theoretical study of the collapse of the harmonically trapped attractively interacting BEC well below the condensate point ($T_c$) started with nonlinear dynamics with the Gross-Pitaevskii equation by Ruprecht \textit{et al} from numerical point \cite{Ruprecht,Pitaevskii} of view and Pitaevskii from analytic point of view \cite{Pitaevskii2} soon after the observation of the Bose-Einstein condensation in the magneto-optical traps \cite{Anderson,Davis,Bradley}. The dynamic properties of such a system in a time-dependent harmonic trap were also analytically studied at this time \cite{Kagan}. Parallelly,  a static theory for the same was proposed with a scaling ansatz by Baym and Pethick \cite{Baym}. Their scaling ansatz results in the critical number of particles for the collapse as $N_c\simeq0.671\frac{l}{a}$ \cite{Dalfovo,Pitaevskii} where $l=\sqrt{\frac{\hbar}{m\omega}}$ is the typical length scales of a single 3-D isotropic harmonic oscillator (\textit{particle of mass $m$ and (angular) frequencies $\omega_x=\omega_y=\omega_z=\omega$})  in the condensate and $a=-a_s>0$ is the magnitude of the $s$-wave scattering length ($a_s<0$) for the attractive inter-particle $g\delta^3_p(\vec{r})$ interactions\footnote{In 3-D, a particle truly is not scattered by another identical particle due to 3-D Dirac delta interaction $V_{int}(\vec{r})=g\delta^3(\vec{r})$ where $g$ is the coupling constant. The scattering in 3-D, however, is possible for a 3-D renormalized Dirac delta interaction ($V_{int}(\vec{r})=g\delta^3(\vec{r})\frac{\partial}{\partial r}\vec{r}$) for which the scattering amplitude takes the form $f(\theta,\phi)=-\frac{mg}{4\pi\hbar^2(1+ikmg/4\pi\hbar^2)}$ for the momentum of incidence $\hbar\vec{k}$ for the elastic scattering. Here, by $g\delta_p^3(\vec{r})$ interaction, we mean the renormalized Dirac delta interaction for a very low ($\hbar\vec{k}$) momentum of incidence. The s-wave scattering length in such a case takes the form $a_s=\frac{mg}{4\pi\hbar^2}$ which becomes negative for $g<0$, i.e., for an attractive interaction. This result is same as the one within the Born approximation for the 3-D Dirac delta interaction $g\delta^3(\vec{r})$ \cite{Pitaevskii1,Bhattacharya}}. Later, the collapse and the subsequent growth/explosion of the condensate were also investigated both theoretically \cite{Sackett,Adhikari} and experimentally \cite{Donley}. For the same system, the critical number of particles was numerically obtained as $N_c=0.575\frac{l}{a}$ \cite{Ruprecht,Dalfovo,Pitaevskii}. The typical length scale, however, is considered to be as $\bar{l}=\big[\frac{\hbar}{m\omega_x}\frac{\hbar}{m\omega_y}\frac{\hbar}{m\omega_z}]^{1/6}$ for an anisotropic trap as well as for an axially symmetric trap ($\omega_z\neq\omega_x=\omega_y=\omega_\perp$) \cite{Dalfovo}.  The critical number of particles was observed to be as $N_c=[0.459\pm0.012\pm0.054]\frac{\bar{l}}{a}$ for $^{85}$Rb atoms (with $a_s\sim-380~a_0$) in a nearly axially symmetric trap ($\omega_x\simeq\omega_y\simeq\sqrt{\omega_x\omega_y}=\omega_\perp$) with  $\frac{\omega_z}{\omega_\perp}=\frac{2\pi\times6.80~\text{Hz}}{2\pi\times17.35~\text{Hz}}\simeq0.39$ \cite{Roberts}. In this experiment, the s-wave scattering length was tuned over a wide range from a positive value to a negative value within the Feshbach resonance \cite{Inouye} to observe a coveralled collapse of the BEC \cite{Roberts}. To observe the controlled collapse, one generally needs to prepare the system in a stable state. Soon after the work of Roberts \textit{et al} \cite{Roberts}, the role of the initial state of the system for the controlled collapse was discussed beyond the Gross-Pitaevskii theory \cite{Kohler}. The collapse of the BEC beyond the mean-field approximation was also theoretically studied after this \cite{Scavage}.  An indication of collapse was also mentioned even during the original observation of the BEC for $^{7}$Li atoms \cite{Bradley}. The critical number of particles for the collapse of the BEC was further obtained numerically for long-ranged interactions which resulted in a better agreement with the experimental data \cite{Haldar,Biswas}. 

The effect of temperature on the collapse had also been studied parallelly from the theoretical point of view. The stability of the same system at a finite temperature was analyzed within the Hartree-Fock approximation \cite{Houbiers}. Temperature dependence in the critical number for the collapse of the BEC was also obtained within the Hartree-Fock-Bogoliubov approximation with the Popov approach \cite{Davis2}. The collapse of the trapped Bose gas was also analytically studied separately for the BEC for $T\rightarrow0$ and the thermal cloud for $T\gg T_c$ within the Hartree-Fock approximation \cite{Mueller}. The temperature dependence of the critical number was later studied by one of us as an extension of the Baym-Pethick scaling theory within the Hartree-Fock approximation for both the condensate and the thermal cloud together for $0\le T<T_c$ \cite{Biswas2}. The applicability of the Hartree-Fock approximation though has a limitation for $T\rightarrow T_c$, it is useful for $0\le T<T_c$ and results the condensate fraction in reasonably good agreement \cite{Giorgini2,Dalfovo,Biswas3} with the experimental data \cite{Ensher}.

The creation of the artificial magnetic field \cite{Lin}, on the other hand, has been drawing a lot of interest for two-photon Raman dressed (electrically neutral) ultracold atoms in a magneto-optical trap \cite{Lin2,Dalibard}. In such a case, a spatially dependent optical coupling between internal states (e.g. $m_F=-1,0,1$) of an ultracold atom, say -- $^{87}$Rb atom (with $a_s\sim109~a_0$ \cite{Dalfovo}) in its $F=1$ hyperfine ground state, results in a Berry phase \cite{Berry} which creates an artificial magnetic field \cite{Lin,Lin2}. Consequently, an electrically neutral atom behaves like a charged particle in a real magnetic field. The Bose-Einstein condensation and many other interesting phenomena such as the superfluid Hall effect \cite{LeBlanc} have been observed in the artificial magnetic field. Landau level phenomena such as the artificial Landau diamagnetism and the artificial de Haas-van Alphen effect \cite{Biswas1} have also been theoretically studied for the neutral atoms in the artificial magnetic field \cite{Biswas4}. However, the collapse of the harmonically trapped Bose gas in artificial magnetic field has not been studied so far. We want to study the same from analytic point of view with a scaling theory for both the BEC and the thermal cloud separately within the Hartree-Fock approximation for the interacting potential of type $g\delta^3_p(\vec{r})$ for any two particles of separation $r$. Essentially, we want to extend both the zero temperature scaling theory and the finite temperature scaling theory for the collapse of the same system exposed to the artificial magnetic field below and above the condensation point. It is expected for a mean-field theory with the Hartree-Fock approximation that the critical exponent ($\alpha$) of the specific heat ($C_v\propto|T-T_c|^{\alpha}$) of the system becomes $0$. This must be true when the system is stable. However, our system is not stable rather metastable. For such a system we would like to study the criticality of the specific heat when the thermal cloud is about to collapse with the critical number of particles ($N=N_c$) just below and above the condensation point. 

The calculations in this article begin in Section-2 with the Hamiltonian for an uncharged particle in a 3-D simple harmonic trap exposed to a uniform artificial magnetic field. Then we write the Gross-Pitaevskii equation for the BEC in such a trap for short ranged attractive inter-particle interactions of the type $g\delta^3_p(\vec{r})$ for $T\rightarrow0$. We describe our scaling ansatz for the collapse of the BEC as an extension of the Baym-Pethick scaling theory for the consideration of the artificial magnetic field. Section-3 starts with the energy functional for the BEC with a perturbative finite temperature effect from the thermal cloud for $0\le T\lnsim T_c$ with the consideration of the Hartree-Fock approximation.  We describe our finite temperature scaling for collapse of the BEC for $0\le T\lnsim T_c$ as an extension of the previous finite temperature scaling theory \cite{Biswas2} for the consideration of the artificial magnetic field. We also describe the collapse of the thermal cloud within the scaling theory for $T\gg T_c$ and $T\simeq T_c$ in Section-4. In this section, we also analyse the criticality of the specific heat ($C_v\propto|T-T_c|^{\alpha}$) when the thermal cloud is about to collapse with the critical number of particles ($N=N_c$) just below and above the condensation point.  Finally, we conclude in Section-5.

\section{Scaling theory for the collapse of the BEC in an artificial magnetic field for $T\rightarrow0$}
Our system of interest is a spin-polarized harmonically trapped Bose gas of uncharged identical (indistinguishable) particles which are exposed to a uniform (constant) artificial magnetic field. In an ideal situation (which is a case of no inter-particle interactions), the single-particle Hamiltonian of the system takes the form \cite{Lin}
\begin{eqnarray}\label{eqn1}
\mathcal{H}(\vec{r},\vec{p})=\bigg[\frac{(\vec{p}-q\vec{A})^2}{2m}+\frac{1}{2}m(\omega_{\perp}^2 r_{\perp}^2 +\omega_z^2 z^2)\bigg]
\end{eqnarray}
where $\vec{p}$ is the momentum and $\vec{r}=\vec{r}_\perp+z\hat{k}$ (with $\vec{r}_\perp=x\hat{i}+y\hat{j}$) is the position of a single particle (boson) of mass $m$ and (angular) frequencies $\omega_x=\omega_y=\omega_\perp$ in the x-y plane and $\omega_z$ along the z-axis in the 3-D simple harmonic trap which is exposed to a uniform (constant) artificial magnetic field $\vec{B}=B\hat{k}$ with the artificial magnetic vector potential $\vec{A}=-\frac{1}{2}\vec{r}\times B\hat{k}=\frac{B}{2}[x\hat{j}-y\hat{i}]$ and the artificial charge $q$. The artificial charge $q$ can either be set as $1$ or be absorbed in $\vec{A}$ in the single-particle Hamiltonian \cite{Stock,Biswas4}. For cylindrical symmetry $\omega_x=\omega_y=\omega_\perp$, the energy eigenvalues of the above single-particle Hamiltonian after the quantization take the form \cite{Halonen,Das,Biswas4}
\begin{eqnarray}\label{eqn1a}
\epsilon_{\tilde{n},\tilde{m},\tilde{j}}&=&(\tilde{n}+1/2)\hbar\big[\sqrt{\omega_\perp^2+\Omega_B^2}+\Omega_B\big]+(\tilde{m}+1/2)\nonumber\\&&\times\hbar\big[\sqrt{\omega_\perp^2+\Omega_B^2}-\Omega_B\big]+(\tilde{j}+1/2)\hbar\omega_z
\end{eqnarray}
where each of the quantum numbers ($\tilde{n},\tilde{m}, \text{and}~\tilde{j}$) takes the values $0,1,2,3,...$ and $\Omega_B=qB/2m$ is half of the cyclotron frequency.  The Bose-Einstein condensation is well described with such a single-particle Hamiltonian in the literature. In the thermodynamic limit ($\bar{\omega}\rightarrow0$, $N\rightarrow\infty$, and $N\bar{\omega}^3=constant$) for $\Omega_B\lesssim\omega_\perp$, the condensation point takes the form \cite{Stock,Biswas4}
\begin{eqnarray}\label{eqn2}
T_c=\frac{\hbar\bar{\omega}}{k_B}\bigg[\frac{N}{\zeta(3)}\bigg]^{1/3}
\end{eqnarray}
where $\bar{\omega}=[\omega_\perp^2\omega_z]^{1/3}$ is the geometric mean of the trap frequencies (which is the same as that of the effective trap frequencies $\sqrt{\omega_\perp^2+\Omega_B^2}+\Omega_B$, $\sqrt{\omega_\perp^2+\Omega_B^2}-\Omega_B$, and $\omega_z$ for the presence of the artificial magnetic field), and $N$ is the total number of particles. Let us now consider the system to be attractively interacting with the inter-particle interactions of the type $g\delta^3_p(\vec{r})$ for any two particles of separation $r$, coupling constant $g=\frac{4\pi\hbar^2a_s}{m}$, and the s-wave scattering length $a_s=-a<0$. The effect of the thermal fluctuations can be ignored well below the condensation point ($T\ll T_c$). For $T\rightarrow0$, the entire system becomes a condensate with a single wave-function $\psi_0(\vec{r},t)$ which follows the Gross-Pitaevskii (G-P) equation
\begin{eqnarray}\label{eqn3}
i\hbar\frac{\partial \Psi_0({\vec{r}},t)}{\partial t}&=& \bigg[-\frac{\hbar^2\nabla^2}{2m}-\Omega_B\hat{L}_z+\frac{m}{2}\big([\omega_\perp^2+\Omega_B^2]r_\perp^2\nonumber\\&&+\omega_z^2z^2\big)+g\mid\Psi_0({\vec{r}},t)\mid^2\bigg]\Psi_0({\vec{r},t})
\end{eqnarray}
where $\psi_0(\vec{r},t)=\psi_0(\vec{r})\text{e}^{-i\mu t/\hbar}$ represents a stationary state with the normalization $\int|\psi_0(\vec{r})|^2\text{d}^3\vec{r}=N_0=N$, $\hat{L}_z=-i\hbar\frac{\partial}{\partial\phi}$ (with $\phi$ as the azimuthal angle) is the z-component of the angular momentum operator, $\mu$ is the chemical potential of the interacting BEC, and $t$ is the time.  In equilibrium for $T\rightarrow0$, the Bose system occupies only the ground state which is same as $\psi_0(\vec{r})$ for the condensate. The normalized ground state in the ideal situation (for no inter-particle interactions) becomes
\begin{eqnarray}\label{eqn3a}
 \psi_{0}^{(0)}(\vec{r})=\sqrt{\frac{N_0}{\pi^{3/2}l_B^2l_z}}\text{e}^{-\frac{r_{\perp}^2}{2l_B^2}} \times\text{e}^{-\frac{z^2}{2l_z^2}}.
\end{eqnarray}	
This is nothing but an eigenstate (i.e. a product of the Fock-Darwin state and the quantum harmonic oscillator state) of the single-particle Hamiltonian in Eqn. (\ref{eqn1}) for $l_B=\sqrt{\frac{\hbar}{m\sqrt{\omega_\perp^2+\Omega_B^2}}}$ \cite{Halonen,Das,Biswas4} and $l_z=\sqrt{\hbar/m\omega_z}$ \cite{Dalfovo}. Now, if we consider inter-particle interactions then according to the scaling ansatz \cite{Baym}, the form of Eqn. (\ref{eqn3a}) primarily remains unaltered, except for a scaling of the length scales $\l_B$ and $l_z$ by a dimensionless factor $\nu$ which is less than $1$ for attractive interactions. Thus, the normalized ground state in the non-ideal situation, as described in  Eqn. (\ref{eqn3}),  becomes \cite{Halonen,Das,Biswas4}
\begin{eqnarray}\label{eqn4}
\psi_{0}(\vec{r})=\sqrt{\frac{N_0}{\pi^{3/2}l_B^2l_z\nu^3}}\text{e}^{-\frac{r_{\perp}^2}{2l_B^2\nu^2}} \times\text{e}^{-\frac{z^2}{2\nu^2l_z^2}}
\end{eqnarray}
according to the scaling theory \cite{Baym,Biswas2} with the scaling parameter $\nu$ which is to be determined by minimizing the stationary-state energy of the system with respect to $\nu$. The stationary-state energy of the BEC for $T\rightarrow0$ can be written from Eqn. (\ref{eqn3}), as
\begin{eqnarray}\label{eqn5}
E_0&=&\int\bigg[\frac{\hbar^2}{2m}\mid\nabla\Psi_0({\vec{r}})\mid^2+\frac{m}{2}\big([\omega_\perp^2+\Omega_B^2]r_\perp^2\nonumber\\&&+\omega_z^2z^2\big)\mid\Psi_0({\vec{r}})\mid^2+\frac{g}{2}\mid\Psi_0({\vec{r}})\mid^4\bigg]\text{d}^3{\vec{r}}.
\end{eqnarray}
Here, $\mid\Psi_0({\vec{r}})\mid^2=n_0(\vec{r})$ is the number density of the BEC at the point $\vec{r}$. It is to be noted that the angular momentum of the BEC (in the ground state) is zero and $\hat{L}_z$ has no contribution to Eqn. (\ref{eqn5}).  The above energy functional can now be recast with the form of $\psi_0(\vec{r})$ of Eqn. (\ref{eqn4}) under the scaling ansatz, as
\begin{eqnarray}\label{eqn6}
E_0(\nu)&=&\frac{N_0\hbar}{2}\bigg\{\sqrt{\omega_\perp^2+\Omega_B^2}+\frac{\omega_z}{2}\bigg\}\Big(\frac{1}{\nu^2}+\nu^2\Big)\nonumber\\&&-\frac{{N}_0^2a}{\sqrt{2\pi}l_B^2l_z}\frac{\hbar^2}{m}\Big(\frac{1}{\nu^3}\Big).
\end{eqnarray} 
The expression of the energy of the BEC, as obtained in Eqn. (\ref{eqn6}), can be written in the dimensionless form ($X_0(\nu)=E_0(\nu)/N_0\hbar\bar{\omega}$) with $N_0=N$, as
\begin{eqnarray}\label{eqn7}
X_0(\nu)&=&\frac{1}{2}\bigg[\frac{\sqrt{\omega_\perp^2+\Omega_B^2}}{\bar{\omega}}+\frac{1}{2}\Big(\frac{\omega_z}{\omega_{\bot}}\Big)^{2/3}\bigg]\Big(\frac{1}{\nu^2}+\nu^2\Big)\nonumber\\&&-\frac{N_0a}{\sqrt{2\pi}l_z}\frac{\sqrt{\omega_{\perp}^2+\Omega_B^2}}{\bar{\omega}}\Big(\frac{1}{\nu^3}\Big).
\end{eqnarray}

\begin{figure}
\includegraphics[width=8.5cm]{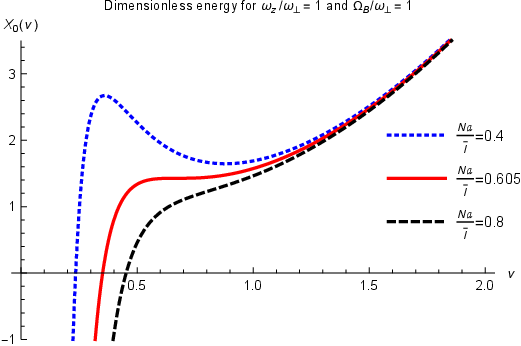}
\caption{The dotted, solid, and dashed lines follow  Eqn. (\ref{eqn7}) for $\frac{Na}{\bar{l}}=0.4, 0.605,~\text{and}~0.8$, respectively with $\frac{\omega_z}{\omega_\perp}=1$ and $\frac{\Omega_B}{\omega_\perp}=1$. 
}
\label{fig1}
\end{figure}

\begin{figure}
\includegraphics[width=8.5cm]{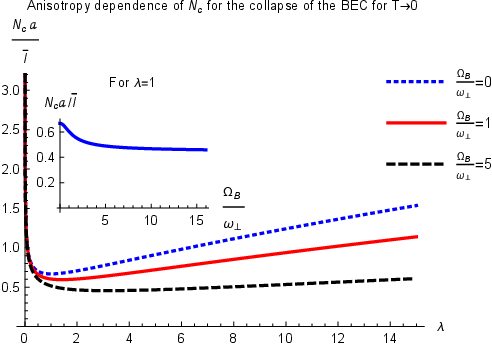}
\caption{The dotted, solid, and dashed lines follow  Eqn. (\ref{eqn8}) for $\frac{\Omega_B}{\omega_\perp}=0, 1,~\text{and}~5$, respectively. The solid line in the inset represents the artificial magnetic field dependence of the same and follows the same equation for $\lambda=1$.
}
\label{fig2}
\end{figure}

We plot the above dimensionless energy with respect to the scaling parameter $\nu$ in figure \ref{fig1} for various fixed values of the number of particles. It is clear from both the figure and the Eqn. (\ref{eqn7}) that energy has a metastable minimum ($\frac{\partial X_0}{\partial\nu}=0$) at a particular $\nu=\nu_c=(1/5)^{1/4}\simeq0.66874$ when the system is critically stable for which we further have to consider $\frac{\partial ^2X_0}{\partial\nu^2}|_{\nu=\nu_c}=0$. It is also clear from both the Eqn. (\ref{eqn7}) and the figure \ref{fig1}  that the system becomes unstable beyond the critical number of particles ($N>N_c$)  such that
\begin{eqnarray}\label{eqn8}
\frac{N_ca}{\bar{l}}&=&\frac{2\sqrt{2\pi}\nu_c}{5}\frac{\frac{1}{\lambda^{1/3}}\big[1+\frac{\lambda}{2}\frac{1}{\sqrt{1+(\Omega_B/\omega_\perp)^2}}\big]}{3/2}\nonumber\\&\simeq&0.670513 \frac{\frac{1}{\lambda^{1/3}}\big[1+\frac{\lambda}{2}\frac{1}{\sqrt{1+(\Omega_B/\omega_\perp)^2}}\big]}{3/2}
\end{eqnarray}
where $\lambda=\frac{\omega_z}{\omega_{\perp}}$ is the anisotropy parameter of the trap. Eqn. (\ref{eqn8}) reveals the anisotropy and the artificial magnetic field dependence of the critical number of particles for the collapse of the BEC. However, we get the original scaling theory result ($\frac{N_ca}{\bar{l}}\simeq0.671$ \cite{Baym,Dalfovo,Pitaevskii}) back from Eqn. (\ref{eqn8}) once we set $\lambda=1$ and $\Omega_B=0$. We plot the above critical number of particles with respect to the anisotropy parameter $\lambda=\frac{\omega_z}{\omega_\perp}$ in figure \ref{fig2} for various fixed values of the artificial magnetic field. We also plot it with respect to the artificial magnetic field for a fixed anisotropy parameter in the inset of the same figure. It is clear from figure \ref{fig2} that the artificial magnetic field causes a reduction of the critical number of particles, such as $\frac{N_ca}{\bar{l}}\simeq0.605$ instead of $0.671$ \cite{Baym,Dalfovo,Pitaevskii} for $\frac{\Omega_B}{\omega_\perp}=1$ and $\lambda=1$, because the magnetic field increases the effective trap frequency for the BEC as indicated in Eqn. (\ref{eqn7}). As a result, the effective length scale of the condensate in the $x-y$ plane gets reduced and consequently, the number density of particles as well as the attractive inter-particle interactions strengthen at around the center of the trap from all the directions in the $x-y$ planes. This causes a reduction in the critical number of particles for the collapse. On the other hand, the anisotropy, i.e. deviation of $\lambda$ from $1$, causes a redistribution of the number density of the particles with a decrement at around the center of the trap. This causes a weakening in the attractive inter-particle intersections at around the center of the trap. Thus it needs more number of particles to collapse the system. It is interesting to note that the critical scaling parameter $\nu_c=(1/5)^{1/4}\simeq0.66874$ \cite{Baym,Dalfovo,Pitaevskii} for which both the 1st order and the 2nd order derivatives of the energy become zero, remains unaltered even for the presence of anisotropy and artificial magnetic field. This is not a surprise because the structure of the scaling variables in the dimensionless energy, such as $(1/\nu^2+\nu^2)$ for the kinetic and potential energies and $1/\nu^3$ for the inter-particle interaction energy, remains unaltered.

It should be mentioned that the scaling result obtained in Eqn. (\ref{eqn8}) is substantially different from the experimental result \cite{Roberts}, even if there is no artificial magnetic field because of the limitations of the scaling theory with the Gross-Pitaevskii equation. The scaling theory ideally requires $\bar{\omega}\rightarrow0$ and $\Omega_B\rightarrow0$ for its applicability. The BEC becomes strongly interacting ($n_0(0)a^3\lesssim1$\footnote{Here, $n_0(0)$ is the number density of particles at the center ($\vec{r}=0$) of the trap.}) close to the collapse. In such a situation, two-body and three-body inelastic collations cause a loss of particles from the condensate \cite{Dalfovo}. These inelastic processes are not captured in the scaling theory. Consideration of these processes along with the finite size effect (for $\bar{\omega}>0$) would substantially improve the scaling result.	 

In a typical experiment for a trapped BEC of $^{87}$Rb atoms in an artificial magnetic field, we have $\omega_z\simeq2\pi\times30$ Hz, $\omega_\perp\simeq2\pi\times95$ Hz, $N\simeq2.5\times10^5$, and $\Omega_B\simeq2\pi\times29$ Hz \cite{Lin2}. In such a set-up of $^{7}$Li atoms (with $a_s\sim-27.4~a_0$ \cite{Roberts}), the critical number for the collapse at $T\rightarrow0$, according to Eqn. (\ref{eqn8}), would be $N_c\simeq2445$. For $^{85}$Rb atoms (with $a_s\sim-380~a_0$ \cite{Roberts}), this critical number would be even less ($N_c\simeq50$). This critical number can be increased by decreasing the absolute value of the s-wave scattering length within the Feshbach resonance \cite{Inouye} even beyond $N_c\sim10000$ \cite{Dalfovo}). However, this critical number further increases if the temperature increases.	 

\section{Finite temperature scaling theory for the collapse of the BEC in an artificial magnetic field}
\begin{figure}
\includegraphics[width=8.5cm]{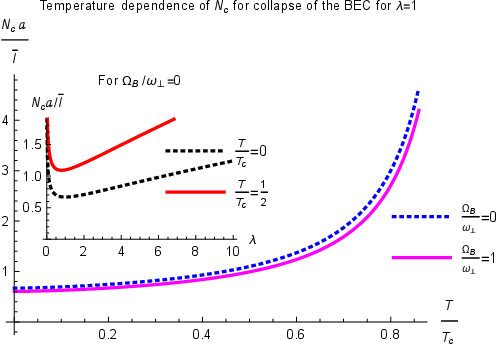}
\caption{The dotted and solid lines follow  Eqn. (\ref{eqn15}) for $\lambda=1$, $a/\bar{l}=0.0066$ \cite{Roberts}, and $\frac{\Omega_B}{\omega_\perp}=0~\text{and}~1$, respectively. The dotted and solid lines in the inset follow the same equation for $\frac{\Omega_B}{\omega_\perp}=0$, $a/\bar{l}=0.0066$ \cite{Roberts}, and $T/T_c=0~\text{and}~1/2$, respectively.
}
\label{fig3}
\end{figure}

At a finite temperature below the condensation point, the BEC is accompanied by the thermal cloud around it.  In an ideal situation, while the average number of particles in the BEC is reduced to $N_0=N(1-[T/T_c]^3)$, the average number of particles in the thermal cloud becomes $N_T=N[T/T_c]^3$ by virtue of the conservation of the total average number of particles $N$. Eventually, the form of the energy of the BEC at a finite temperature $T$ below the condensation point ($T<T_c$) is modified from Eqn. (\ref{eqn5}) to the following \cite{Giorgini,Pitaevskii3}
\begin{eqnarray}\label{eqn9}
E_0&=&\int\bigg[\frac{\hbar^2}{2m}\mid\nabla\Psi_0({\vec{r}})\mid^2+\frac{m}{2}\big([\omega_\perp^2+\Omega_B^2]r_\perp^2+\omega_z^2z^2\big)\nonumber\\&&\times\mid\Psi_0({\vec{r}})\mid^2+\frac{g}{2}\big[\mid\Psi_0({\vec{r}})\mid^2+4n_T(\vec{r})\big]\mid\Psi_0({\vec{r}})\mid^2\bigg]\text{d}^3{\vec{r}}\nonumber\\
\end{eqnarray}
where $n_T(\vec{r})$ is the number density of the thermal cloud over the BEC at a temperature $T$ at the point $\vec{r}$. Eqn. (\ref{eqn9}) is a result within the Hartree-Fock approximation \cite{Pitaevskii3}. The number density in the case of no artificial magnetic field takes the following form \cite{Giorgini2,Pitaevskii3} 
\begin{eqnarray}\label{eqn10}
n_T(\vec{r})=\frac{1}{\lambda_T^3}\text{Li}_{3/2}(\text{e}^{-\big[\frac{m}{2}[\omega_\perp^2r_\perp^2+\omega_z^2z^2]+2gn(\vec{r})-2gn(0)\big]/k_BT})\nonumber\\
\end{eqnarray}
within the Hartree-Fock approximation with the total number density of particles $n(\vec{r})=n_0(\vec{r})+n_T(\vec{r})$. Here, $\text{Li}_j(u)=u+u^2/2^j+u^3/3^j+...$ is a poly-logarithmic function of order $j=3/2$ and argument $u=\text{e}^{-\big[\frac{m}{2}[\omega_\perp^2r_\perp^2+\omega_z^2z^2]+2gn(\vec{r})-2gn(0)\big]/k_BT}$ and $\lambda_T=\sqrt{\frac{2\pi\hbar^2}{mk_BT}}$ is the thermal de Broglie wavelength. In an ideal case, the number density $n_T(\vec{r})$ of the thermal cloud is obtained by integrating the average number of particles ($n_T(\vec{r})=\int\frac{1}{\text{e}^{\mathcal{H}(\vec{r},\vec{p})/k_BT}-1}\frac{\text{d}^3\vec{p}}{(2\pi\hbar)^3}$ according to Bose-Einstein statistics for $0<T<T_c$) over the momentum. The $\vec{p}-q\vec{A}$ term in the single-particle Hamiltonian ($\mathcal{H}(\vec{r},\vec{p})$) though is a shift in momentum, the artificial magnetic field contributes nothing to the number density of the thermal cloud after integrating the Gaussian functions (like $\text{e}^{-(\vec{p}-q\vec{A})^2/2mk_BT}$) over the momentum. Thus the number density of the thermal cloud takes the following form
\begin{eqnarray}\label{eqn11}
n_T(\vec{r})=\frac{1}{\lambda_T^3}\text{Li}_{3/2}\big(\text{e}^{-\big[\frac{m}{2}[\omega_\perp^2r_\perp^2+\omega_z^2z^2]\big]/k_BT}\big)
\end{eqnarray}
even for the case of the presence of the artificial magnetic field in an ideal situation ($g=0$). According to the finite temperature scaling theory, the effects of the inter-particle interactions on the density of the thermal cloud can be accounted with the scaling parameter $\nu$ in the following form \cite{Biswas2}
\begin{eqnarray}\label{eqn12}
n_T(\vec{r})=\frac{1}{\nu^3\lambda_T^3}\text{Li}_{3/2}\big(\text{e}^{-\big[\frac{m}{2}[\omega_\perp^2r_\perp^2+\omega_z^2z^2]\big]/\nu^2k_BT}\big)
\end{eqnarray}
for $T\le T_c$. It is interesting to note that the expressions of the condensation point $T_c=\frac{\hbar\bar{\omega}}{k_B}\big(\frac{N}{\zeta(3)}\big)^{1/3}$, total average number of particles ($N_T=\int n_T(\vec{r})\text{d}^3\vec{r}=N[T/T_c]^{3}$) in the thermal cloud, and total average number of particles ($N_0=N-N_T=N\big(1-[T/T_c]^{3}\big)=\int|\psi_{0}(\vec{r})|^2\text{d}^3\vec{r}$) in the BEC  remain unaltered under such a scaling \cite{Biswas2}. For any thermodynamic system of a fixed (average) total number of particles at a finite temperature, the system equilibrates through the minimization of the Helmholtz free energy ($F=E-TS$). The (average) entropy ($S$\footnote{Here, the (average) entropy of the trapped system takes the form $S=Nk_B\big[4(T/T_c)^3\frac{\text{Li}_4(\text{e}^{\mu/k_BT})}{\zeta(3)}-\frac{\mu}{k_BT}\big]$ where $\mu=\bar{\mu}+\hbar[\sqrt{\omega_\perp^2+\Omega_B^2}+\omega_z/2]$ is the chemical potential of the system with $\bar{\mu}<0$ for $T>T_c$ and $\mu-\bar{\mu}\rightarrow0$ in the thermodynamic limit. For $T\le T_c$, on the other hand, the total (average) number of particles ($N$) is to be replaced with the total (average) number of particles in the thermal cloud ($N_T$) with $\bar{\mu}=0$ in the formula of the (average) entropy.}), however, remains unaltered under the scaling because the entropy per unit volume is independent of the number density \cite{Pethick}. Hence only the minimization of the energy would be relevant for the scaling theory. Thus we recast Eqn. (\ref{eqn9}) within the finite temperature scaling theory for $\hbar\Omega_B\lesssim\hbar\bar{\omega}\ll k_BT<k_BT_c$ with Eqns. (\ref{eqn4}) and (\ref{eqn12}), as
\begin{eqnarray}\label{eqn13}
E_0(\nu)&\simeq&\frac{N_0\hbar}{2}\bigg\{\sqrt{\omega_\perp^2+\Omega_B^2}+\frac{\omega_z}{2}\bigg\}\Big(\frac{1}{\nu^2}+\nu^2\Big)\nonumber\\&&-\frac{N_0^2a}{\sqrt{2\pi}l_B^2l_z}\frac{\hbar^2}{m}\bigg[1+\frac{4}{N_0}\Big[\frac{N_T}{\zeta(3)}\Big]^{1/2}\frac{\zeta(3/2)}{\sqrt{1+\Omega_B^2/\omega_{\perp}^2}}\bigg]\nonumber\\&&\times\Big(\frac{1}{\nu^3}\Big).
\end{eqnarray} 
Now the above expression for energy can be recast in dimensionless form ($X_0(\nu)=E_0(\nu)/N\hbar\bar{\omega}$) in terms of the temperature and the scaling parameter, as
\begin{eqnarray}\label{eqn14}
X_0(\nu)&\simeq&\frac{1-[T/T_c]^3}{2}\bigg[\frac{\sqrt{\omega_\perp^2+\Omega_B^2}}{\bar{\omega}}+\frac{1}{2}\Big(\frac{\omega_z}{\omega_{\bot}}\Big)^{2/3}\bigg]\Big(\frac{1}{\nu^2}\nonumber\\&&+\nu^2\Big)-\frac{Na(1-[T/T_c]^3)^2}{\sqrt{2\pi}l_z}\frac{\sqrt{\omega_{\perp}^2+\Omega_B^2}}{\bar{\omega}}\bigg[1\nonumber\\&&+\frac{4}{N^{1/2}}\frac{[T/T_c]^{3/2}}{1-[T/T_c]^3}\frac{\zeta(3/2)/\zeta^{1/2}(3)}{\sqrt{1+\Omega_B^2/\omega_{\perp}^2}}\bigg]\Big(\frac{1}{\nu^3}\Big)~~~~~~
\end{eqnarray}
where the structure of the scaling function remains unaltered even after incorporating finite temperature effects. Consequently, the BEC will be unstable beyond the critical number of particles ($N>N_c$) in the system (BEC+thermal cloud). Hence we follow the same procedure, as described in the previous section, to determine the critical number of particles for the collapse of the BEC at a finite temperature. It is interesting to note that 2nd interaction term in Eqn. (\ref{eqn14}) is a perturbation to the 1st interaction term for $a/\bar{l}\ll1$ \cite{Roberts} and  $0<T\lnsim T_c$. By following the same procedure, from Eqn. (\ref{eqn14}) we obtain the critical number of particles ($N_c$) for a low temperature specially for $0<T\lnsim T_c$, as
\begin{eqnarray}\label{eqn15}
\frac{N_ca}{\bar{l}}&\simeq&0.670513 \frac{\frac{1}{\lambda^{1/3}}\big[1+\frac{\lambda}{2}\frac{1}{\sqrt{1+(\Omega_B/\omega_\perp)^2}}\big]}{3/2}\frac{1}{1-[T/T_c]^3}\Bigg[1\nonumber\\&&+14.255254 \lambda^{1/6}\frac{1}{\sqrt{1+\frac{\lambda}{2}\frac{1}{\sqrt{1+\Omega_B^2/\omega_\perp^2}}}}\frac{[T/T_c]^{3/2}}{\sqrt{1-[T/T_c]^3}}\nonumber\\&&\times\bigg(\frac{a}{\bar{l}}\bigg)^{1/2}+\mathcal{O}([a/\bar{l}]^{1})\Bigg].
\end{eqnarray}
We plot Eqn. (\ref{eqn15}) to show the temperature dependence of the critical number of particles in the Bose system required for the collapse of the BEC in figure \ref{fig3} for various values of the artificial magnetic field. It is interesting to note that the critical number increases with temperature because the number of particles in the BEC decreases with the temperature requiring more particles in contributing to the inward pressure due to the attractive interactions in the BEC over the outward quantum (for $T\rightarrow0$) and thermal (for $T\neq0$) pressure of the non-interacting parts of the BEC. The artificial magnetic field further reduces the outward pressure as discussed in the previous section. So, it reduces the critical number of particles. We also plot Eqn. (\ref{eqn15}) to show the anisotropy dependence of the critical number of particles for various temperatures in the inset of the same figure. It is interesting to note that the effect of the anisotropy is stronger at higher temperature in contrary to the effect of anisotropy on the bulk properties of a thermodynamic system. This is actually not a surprise because the critical number is not an extensive variable in the trap ($N_c[\hbar\bar{\omega}]^3\rightarrow0$) at least for $0\le T\lnsim T_c$.

For $T\rightarrow T_c$, there exists no BEC. Hence there is no question of the collapse of the BEC for $T\rightarrow T_c$. However, the thermal cloud can collapse near the condensation point. This will allow us to analyse the critically of the critical number of collapse of the thermal cloud.

\section{Finite temperature scaling theory for criticality at the critical number of particles}
The collapse is a property of the attractive inter-particle interactions which are also there among the particles in the thermal cloud. The number of particles required to collapse the thermal cloud must be much larger \cite{Mueller,Biswas2} than that required for the collapse of a BEC. Thus only the BEC can collapse if the required number of particles for the collapse of the thermal cloud is not met. However, there is no BEC for $T\ge T_c$. Only the thermal cloud can collapse for this range of temperature. The study of the same leads to a critically at the critical number for the collapse of the thermal cloud near around $T=T_c$. Let us also study the same within the domain of the scaling theory for the same system of our interest.

\subsection{Scaling theory for the collapse of the thermal cloud for $T\gtrsim T_c$ and $\Omega_B\lesssim\omega_\perp$}
For the case of attractive interactions, the average number density of particles can be obtained within the scaling ansatz by multiplying $(\vec{p}-q\vec{A})^2/2m$ with $\nu^2$ and $\frac{m}{2}[\omega_\perp^2 r_\perp^2+\omega_z^2z^2]$ by $1/\nu^2$ in the expression of the average number of particles. The purpose of proposing such a scaling ansatz is to reduce the typical length scale of the harmonic oscillator by the factor $\nu$ ($\nu<1$). Eventually, the average potential energy decreases by the factor $\nu^2$, kinetic energy increases by the factor $1/\nu^2$, and the average attractive inter-particle interaction energy decreases by the factor $1/\nu^3$  keeping the total average number of particles and $T_c$ unaltered as already described in the previous section.  Thus semi-classically, from Eqn. (\ref{eqn1}) we obtain the average number density of particles for the interacting system within the scaling ansatz on the Bose-Einstein statistics for $T>T_c$, as
\begin{eqnarray}\label{eqn16}
n_T(\vec{r})&=&\int\frac{1}{\text{e}^{\big[\frac{(\vec{p}-q\vec{A})^2\nu^2}{2m}+\frac{m[\omega_\perp^2 r_\perp^2+\omega_z^2z^2]}{2\nu^2}-\mu\big]/k_BT}-1}\frac{\text{d}^3\vec{p}}{(2\pi\hbar)^3}\nonumber\\&=&\frac{1}{\nu^3\lambda_T^3}\text{Li}_{3/2}\big(\text{e}^{-\big[\frac{m}{2\nu^2}[\omega_\perp^2r_\perp^2+\omega_z^2z^2]-\mu\big]/k_BT}\big)
\end{eqnarray}
where $\mu=\bar{\mu}+\hbar[\sqrt{\omega_\perp^2+\Omega_B^2}+\omega_z/2]$ is the chemical potential of the system such that the 1st term $\bar{\mu}$ ($\le0$) represents the chemical potential (which essentially is the chemical potential of a 3-D harmonically trapped ideal Bose gas) in the thermodynamic limit and the second term $\hbar[\sqrt{\omega_\perp^2+\Omega_B^2}+\omega_z/2]$ represents the ground state energy of the system for $g=0$.  The total average energy of the interacting system $E=\int\big[(\vec{p}-q\vec{A})^2/2m+\frac{m}{2}[\omega_\perp^2 r_\perp^2+\omega_z^2z^2]+gn_T(\vec{r})\big]n_T(\vec{r})\text{d}^3\vec{r}$ can now be obtained within the Hartree-Fock approximation \cite{Pitaevskii} and the finite temperature scaling ansatz as described in Eqn. (\ref{eqn16}) for $T>T_c$, as
\begin{eqnarray}\label{eqn17}
E_>&=&\bigg[Nk_BT\frac{3}{2}\frac{\text{Li}_4(\text{e}^{\bar{\mu}/k_BT})}{\text{Li}_3(\text{e}^{\bar{\mu}/k_BT})}+\frac{3}{4}N\hbar[\sqrt{\omega_\perp^2+\Omega_B^2}+\omega_z/2]\bigg]\nonumber\\&&\times\Big(\frac{1}{\nu^2}+\nu^2\Big)-N^{3/2}\hbar\bar{\omega}\sqrt{\frac{2}{\pi}}\frac{1}{\zeta^{3/2}(3)}\frac{a}{\bar{l}}\bigg(\frac{T}{T_c}\bigg)^{9/2}\frac{1}{\nu^3}\nonumber\\&&\times\sum_{i=1}^{\infty}\sum_{j=1}^{\infty}\frac{\text{e}^{(i+j)\bar{\mu}/k_BT}}{i^{3/2}j^{3/2}[i+j]^{3/2}}
\end{eqnarray}
where
\begin{eqnarray}\label{eqn18}
N=\big(\frac{k_BT}{\hbar\bar{\omega}}\big)^3\text{Li}_3(\text{e}^{\bar{\mu}/k_BT})
\end{eqnarray}
is the total average number of particles in the thermodynamic limit for $\Omega_B\lesssim\omega_\perp$. On the other hand, we have $\bar{\mu}=0$ and $N_T=N(T/T_c)^3$ for the thermal cloud for $T\le T_c$. Thus from Eqn. (\ref{eqn17}), we obtain the average total energy of the thermal cloud  for $T\le T_c$, as
\begin{eqnarray}\label{eqn17b}
E_<&=&\bigg[Nk_BT\bigg(\frac{T}{T_c}\bigg)^{3}\frac{3}{2}\frac{\zeta(4)}{\zeta(3)}+\frac{3}{4}N\hbar[\sqrt{\omega_\perp^2+\Omega_B^2}+\omega_z/2]\bigg]\nonumber\\&&\times\Big(\frac{1}{\nu^2}+\nu^2\Big)-N^{3/2}\hbar\bar{\omega}\sqrt{\frac{2}{\pi}}\frac{1}{\zeta^{3/2}(3)}\frac{a}{\bar{l}}\bigg(\frac{T}{T_c}\bigg)^{9}\frac{1}{\nu^3}\nonumber\\&&\times\sum_{i=1}^{\infty}\sum_{j=1}^{\infty}\frac{1}{i^{3/2}j^{3/2}[i+j]^{3/2}}.
\end{eqnarray}
The magnetic field dependent part in both Eqn. (\ref{eqn17}) and Eqn. (\ref{eqn17b}) essentially becomes negligible in the thermodynamic limit for $\Omega_B\lesssim\omega_\perp$ according to Bohr-van Leeuwen theorem. We are considering such a case for the rest of the discussions.

\subsubsection{The collapse for $T\gg T_c$}
The chemical potential in the thermodynamic limit for $\Omega_B\lesssim\bar{\omega}$ is to be determined from Eqn. (\ref{eqn18}) and it is well known in the literature \cite{Biswas5}. In such a case for $T\gg T_c$, we have $\bar{\mu}\simeq-k_BT\ln\big([\frac{T}{T_c}]^3/\zeta(3)\big)$. Thus we obtain the total average energy of the system after the finite temperature scaling for $T\gg T_c$, as
\begin{eqnarray}\label{eqn19}
E_>&\simeq&N^{4/3}\hbar\bar{\omega}\frac{T}{T_c}\frac{1}{\zeta^{1/3}(3)}\frac{3}{2}\Big(\frac{1}{\nu^2}+\nu^2\Big)-N^{3/2}\hbar\bar{\omega}\sqrt{\frac{2}{\pi}}\frac{1}{\zeta^{3/2}(3)}\nonumber\\&&\times\frac{a}{\bar{l}}\bigg(\frac{T}{T_c}\bigg)^{-3/2}\frac{\zeta^2(3)}{2^{3/2}}\frac{1}{\nu^3}.
\end{eqnarray}
It is interesting to note from the above that the absolute value of the inter-particle interaction energy decreases as $1/T^{3/2}$ for $T\gg T_c$. From Eqn. (\ref{eqn19}) we obtain the critical number of particles for the collapse of the thermal cloud by following the method already described in Section-2, as
\begin{eqnarray}\label{eqn20}
N_c\simeq18.539131\bigg[\frac{\bar{l}}{a}\bigg]^6\bigg[\frac{T}{T_c}\bigg]^{15}.
\end{eqnarray}
Eqn. (\ref{eqn20}) is, of course, a classical result for the collapse and consequently,  a similar result with a replacement of the condensation point $T_c$ by the Fermi temperature $T_F=\frac{1}{k_B}\hbar\bar{\omega}(6N/s)^{1/3}$ \cite{Giorgini,Biswas5} is also applicable for attractively interacting fermions in the harmonic trap. It is interesting to note that for the Bose system, the exponent of $\bar{l}/a$ in the critical number ($N_c$) changes from $1$ to $6$, with an intermediate exponent $3/2$ for $0 \lnsim T/T_c\lnsim 1$\footnote{See Eqn. (\ref{eqn15}).}, as the temperature changes from $0$ to a very high ($T\gg T_c$) value. Interestingly, the exponent of $T/T_c$ in $N_c$ becomes $15$ for $T\gg T_c$. Because of such a large exponent, the observation of the collapse would require a much greater number of particles in comparison to that required for $T\rightarrow0$. This exponent, however, will be changed if the temperature is lowered towards $T_c$. Let us now analyse the criticality for the collapse near around the critical (condensation) point.

\subsubsection{The collapse at around $T=T_c$}
Just above the condensation point ($T\gtrapprox T_c$) the chemical potential takes the form $\mu\simeq-3k_BT_c\frac{\zeta(3)}{\zeta(2)}[T/T_c-1]$ \cite{Biswas6} from Eqn. (\ref{eqn18}). Now to the first order in $\bar{\mu}\simeq-3k_BT_c\frac{\zeta(3)}{\zeta(2)}[T/T_c-1]$ for $\Omega_B\lesssim\omega_{\perp}$, we recast Eqn. (\ref{eqn17}) as
\begin{eqnarray}\label{eqn21}
E_>&\simeq&Nk_BT\bigg[\frac{3}{2}\frac{\zeta(4)}{\zeta(3)}-\frac{9}{2}\frac{\zeta(3)}{\zeta(2)}\bigg(1-\frac{\zeta(4)\zeta(2)}{\zeta^2(3)}\bigg)[1-T_c/T]\bigg]\nonumber\\&&\times\Big(\frac{1}{\nu^2}+\nu^2\Big)-N^{3/2}\hbar\bar{\omega}\sqrt{\frac{2}{\pi}}\frac{1}{\zeta^{3/2}(3)}\frac{a}{\bar{l}}\bigg(\frac{T}{T_c}\bigg)^{9/2}\frac{1}{\nu^3}\nonumber\\&&\times\bigg[0.664717-2.416942\frac{3\zeta(3)}{\zeta(2)}[1-T_c/T]\bigg]
\end{eqnarray}
in the thermodynamic limit. For $T\le T_c$ on the other hand, $N$ will be replaced with $N(T/T_c)^3$ and $\bar{\mu}=0$ for the thermal cloud. Thus energy of the thermal cloud after the scaling takes the from Eqn. (\ref{eqn17b}), as
\begin{eqnarray}\label{eqn22}
E_<&\simeq&Nk_BT\bigg(\frac{T}{T_c}\bigg)^{3}\bigg[\frac{3}{2}\frac{\zeta(4)}{\zeta(3)}\Big(\frac{1}{\nu^2}+\nu^2\Big)\bigg]-N^{3/2}\hbar\bar{\omega}\sqrt{\frac{2}{\pi}}\nonumber\\&&\times\frac{1}{\zeta^{3/2}(3)}\frac{a}{\bar{l}}\bigg(\frac{T}{T_c}\bigg)^{9}\frac{0.664717}{\nu^3}
\end{eqnarray}
for $T\lessapprox T_c$. 

The critical number of particles at the condensation point ($T=T_c$) can be calculated from either of the above two equations by following the same procedure as described in Section-2, as
\begin{eqnarray}\label{eqn23}
N_c\simeq2.035592\bigg[\frac{\bar{l}}{a}\bigg]^6.
\end{eqnarray}
This result is close to the one obtained in Ref. \cite{Biswas2}. A difference in the decimal fractions of the two results is coming from the approximation involved in the numerical summation of $\sum_{i=1}^{\infty}\sum_{j=1}^{\infty}1/[i^{3/2}j^{3/2}(i+j)^{3/2}]\simeq0.664717$. The collapse at around the condensation point can be further studied by considering Eqns. (\ref{eqn21}) and (\ref{eqn22}) separately for $T\gtrapprox T_c$ and $T\lessapprox T_c$, respectively. Eqns. (\ref{eqn21}) and (\ref{eqn22}) can be useful to the study the critically of the specific heat of the system when the system is about to collapse.

In a typical experiment, we have $\bar{l}/a\sim0.0064$ \cite{Roberts}. For such a case, the critical number at $T=T_c$ according to  Eqn. (\ref{eqn23}), becomes $N_c\sim3\times10^{13}$. This number can be reduced to $N_c\sim5\times10^{5}$ by increasing the absolute value of the scattering length 20 times within the Feshbach resonance \cite{Inouye}. In this situation, the critical number at $T/T_c=10$ would be $N_c\sim4\times10^{21}$ according to  Eqn. (\ref{eqn20}). Accommodation of such a large number of particles in a magneto-optical trap has not been reported so far.

\subsection{Critically of the specific heat at the critical number of particles}
\begin{figure}
\includegraphics[width=8.5cm]{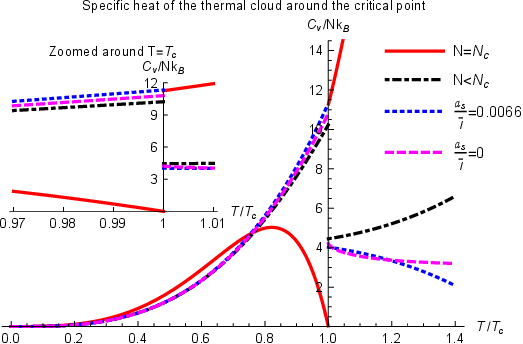}
\caption{Both the dot-dashed and the solid lines follow Eqn. (\ref{eqn25}) for the set of parameters $\{N=10^8,~\frac{a_s}{\bar{l}}=-0.0066,~\nu=0.982341\}$\cite{Num} and $\{N=N_c~\text{for}~T=T_c,~\nu=\nu_c\}$, respectively. Both the dotted and the dashed lines follow Eqn. (\ref{eqn24}) with $E_<$ and $E_>$ from Eqns. (\ref{eqn17b}) and (\ref{eqn17}) for the set of parameters $\{N=10^8,~\frac{a_s}{\bar{l}}=0.0066,~\nu=1.01622\}$\cite{Num2} and $\{\frac{a_s}{\bar{l}}=0\}$, respectively. The inset represents a zoom of the same at around $T=T_c$.
}
\label{fig4}
\end{figure}

The BEC though contributes to the energy, it does not contribute to the specific heat because there are no energy fluctuations in the condensate. Only thermal cloud contributes to the specific heat. The specific heat\footnote{Here, $\theta(T-T_c)$ is a unit step function.}
\begin{eqnarray}\label{eqn24}
C_v=\frac{\partial E_<}{\partial T}\bigg|_{\bar{\omega},N}\theta(T_c-T)+\frac{\partial E_>}{\partial T}\bigg|_{\bar{\omega},N}\theta(T-T_c)
\end{eqnarray}
of the system at constant geometric mean frequency ($\sim$ volume) and number of particles in the system can be obtained, as
\begin{eqnarray}\label{eqn25}
C_v&\simeq&Nk_B\bigg[5.402356\bigg(\frac{1}{\nu^2}+\nu^2\bigg)\big(\frac{T}{T_c}\big)^{3}-3.851\frac{aN^{1/6}}{\bar{l}\nu^3}\big(\frac{T}{T_c}\big)^{8}\bigg]\nonumber\\&&\times\theta(T_c-T)+Nk_B\bigg[2.113923\bigg(\frac{1}{\nu^2}+\nu^2\bigg)+\frac{aN^{1/6}}{\bar{l}\nu^3}\nonumber\\&&\times\big[13.423169 \big(\frac{T}{T_c}\big)^{7/2}-11.937854\big(\frac{T}{T_c}\big)^{5/2}\big]\bigg]\theta(T-T_c)\nonumber\\
\end{eqnarray}
once the expressions for the energies $E_<$ and $E_>$ are taken from Eqns. (\ref{eqn22}) and  (\ref{eqn21}), respectively. Eqn. (\ref{eqn25}) is the specific heat of the system at around the critical point $T_c$. The critical exponent ($C_v\propto|T-T_c|^{\alpha}$) of the specific heat\footnote{$C_v=Nk_B\frac{1}{\bar{l}\nu^3}\big[3.203675 (1.6863\bar{l}\nu+1.6863\bar{l}\nu^5 - 1.202057 a N^{1/6})+25.6294 (0.632362\bar{l}\nu + 0.632362\bar{l}\nu^5 - 1.20206 a N^{1/6})\frac{T-T_c}{Tc}+\mathcal{O}([T-T_c]^2)\big]$ for $T\lessapprox T_c$ and $C_v=Nk_B\frac{1}{\bar{l}\nu^3}\big[0.125197 (16.884771\bar{l}\nu+16.884771\bar{l}\nu^5 + 11.863826 a N^{1/6})+17.136458 a N^{1/6}\frac{T-T_c}{Tc}+\mathcal{O}([T-T_c]^2)\big]$ for $T\gtrapprox T_c$.} for $T$  very close to $T_c$, however, takes the value $\alpha=0$ whenever the system is stable ($a_s>0$) or metastable ($a_s<0$) below the critical number of particle ($N<N_c$) in-spite of having a discontinuity at $T=T_c$ and different slopes at around $T=T_c$. This is an expected result for a mean-field theory with the Hartree-Fock approximation. However, interestingly, the critical exponent $\alpha$ dramatically changes to $\alpha=1$ for $T-T_c=0_-$ once the critical number of particles ($N=N_c$) with the critical scaling parameter $\nu=\nu_c$ and the condition Eqn. (\ref{eqn23}) are taken to meet the critical collapse at $T={T_c}_{-}$. The specific heat in such a situation takes the form
\begin{eqnarray}\label{eqn26}
C_v=-72.480212Nk_B\bigg[\frac{T-T_c}{T_c}\bigg]^{1}
\end{eqnarray}
for $T-T_c=0_-$. The critical exponent $\alpha$, of course, takes the value $0$ for the case of the critical collapse ($N=N_c$) at $T={T_c}_+$. The dramatic behaviour of the critical exponent $\alpha$ is not a surprise. Actually, below $T_c$, the (average) number of particles in the thermal cloud decreases, and that in the condensate increases if the temperature is lowered. Consequently, the inward pressure due to the attractive interactions in the thermal cloud is reduced more in comparison to the reduction of the outward thermal pressure in the thermal cloud. This does not allow the thermal cloud to collapse for $T\le T_c$ except at $T_c$ even if $N=N_c$ of $T=T_c$ is taken. At $T_c$, there is no condensate to reduce the inward pressure of the thermal cloud. The thermal cloud collapses if $N=N_c$ is taken at $T=T_c$ according to the criterion as mentioned in Eqn. (\ref{eqn23}). However, if the temperature is increased beyond $T_c$ with $N=N_c$ of $T=T_c$, then the effective volume of the system increases along-with the thermal energy resulting in a decrease in inward pressure due to the attractive interactions and an increase in outward thermal pressure. This does not allow the thermal cloud to collapse for $T\ge T_c$ even if $N=N_c$ of $T=T_c$ is taken. In addition to the above, the specific heat ($C_v$) is discontinuous for the trapped system at $T=T_c$ because its chemical potential ($\mu$) although continuous, is non-analytic at $T=T_c$.  For all these reasons, we have a dramatic change in the critical exponent of the specific heat ($C_v\propto|T-T_c|^{\alpha}$) from $\alpha=1$ to $\alpha=0$ if the temperature increases across $T=T_c$ with $N=N_c$ of $T=T_c$.

We plot Eqn. (\ref{eqn25}) in figure \ref{fig4} for two cases of $N<N_c$ (dot-dashed line) and $N=N_c$ (solid line) with respect to temperature. We also plot Eqn. (\ref{eqn24}) in figure \ref{fig4} for two cases of repulsive interactions (dotted line) and no interactions (dashed line). It is also clear from the plots that $\alpha=0$ as long as $N<N_c$. It is also clear from figure \ref{fig4} that while the repulsive (attractive) interactions cause an increase in the specific heat for $T<T_c$, the repulsive (attractive) interactions cause a decrease in the specific heat for $T>T_c$. Such a behaviour of the specific heat for various number of particles near the critical number $N_c$ was indicated numerically before us \cite{Goswami}. The dramatic change in the critical exponent from $1$ to $0$ for the critical collapse ($N=N_c$) at $T-T_c=0_-$ and $T-T_c=0_+$, respectively, is also apparent in figure \ref{fig4}. Interestingly, the expression for the specific heat in Eqn. (\ref{eqn26}) becomes scale free (i.e. independent of the s-wave scattering length $a_s $ and the typical length scale $\bar{l}$ of the trap) at $T-T_c=0_-$ because of the use of Eqn. (\ref{eqn23}).

\section{Conclusion}
To conclude, we have analytically explored both the zero temperature and the finite temperature scaling theory for short ranged (contact) attractive inter-particle interactions within the Hartree-Fock approximation for the collapse of an attractively interacting 3-D harmonically trapped Bose gas in an artificial magnetic field. We have separately studied the collapse of both the condensate and the thermal cloud below and above the condensation point, respectively. We have obtained an anisotropy, artificial magnetic field, and temperature dependent critical number of particles for the collapse of the condensate. We also have obtained a temperature dependent critical number of particles for the collapse of the thermal cloud in both the classical ($T\gg T_c$) regime and the critical ($T\simeq T_c$) regime with an emphasis on the criticality at the critical number of particles.

Eqns. (\ref{eqn8}), (\ref{eqn15}), (\ref{eqn20}), (\ref{eqn23}), (\ref{eqn25}), and (\ref{eqn26}) are the key results obtained by us. While the effect of the artificial magnetic field, as shown in figure \ref{fig3}, becomes important in decreasing the critical number of particles for the collapse, the effect of anisotropy of the trap, as shown in figure \ref{fig2}, becomes important in increasing the critical number. The effect of increasing the temperature, as shown in figure \ref{fig3}, becomes important in further increasing the critical number of particles for the collapse. The criticality of the specific heat ($C_v\propto|T-T_c|^{\alpha}$) of the system (thermal cloud) at the critical number of particles ($N=N_c$), as shown in figure \ref{fig4}, becomes very interesting as it shows a dramatic change (from $\alpha=1$ to $0$) in the critical exponent $\alpha$ across the condensation point $T_c$. More interestingly, the specific heat ($C_v$) just below $T_c$ for $N=N_c$ becomes scale free. All these results, except Eqn. (\ref{eqn20}), are experimentally testable within the present day experimental set-up for the ultracold systems in the magneto-optical traps. It will be a challenge for experimentalists to improve their magneto-optical trap to accommodate about $10^{21}$ (or more) atoms in order to test our Eqn. (\ref{eqn20}).

The finite temperature scaling theory within the Hartree-Fock approximation though explains the basic physics of the collapse, it is not free from drawbacks, e.g. very close to the critical point of a stable system, the mean field theory fails to predict a nonzero value of the critical exponent in contrary to the experimental data $\alpha=-0.0127\pm0.0003$ \cite{Lipa} for the $\lambda$ transition of liquid $^4$He\footnote{In reality, generally the specific heat at constant pressure ($C_p$) is measured for convenience and the $C_v$ is obtained from the thermodynamic relation between the $C_p$ and $C_v$. The measured value of the critical exponent for the $C_p$ at around the $\lambda$ point is $\alpha=-0.0127\pm0.0003$. The critical exponents ($\alpha$) for both the $C_p$ and the $C_v$ are the same for the $\lambda$ transition \cite{Annett}.}. On the other hand, for a repulsively interacting harmonically trapped Bose gas, the renormalization group result for the trap critical exponent ($\theta$) for the correlation length ($\xi=\bar{l}^\theta$) at $T=T_c$ is $\theta=0.57327(4)$ within the 3-D $X-Y$ model \cite{Campostrini}. Away from $T_c$, the correlation length takes the untapped form $\xi=\xi_o|1-T/T_c|^{-\nu}$ with the experimental value $\nu=0.67\pm0.13$ or $\alpha=2-3\nu=-0.01\mp0.39$ \cite{Donner} if $\xi\ll\bar{l}^\theta$ \cite{Campostrini}. The finite temperature scaling theory does not also take the entropy of the system into account as it remains unaltered under the scaling. We have obtained a nonzero value of $\alpha$ just below $T_c$ for $N=N_c$ within a mean-field theory because the system is at the criticality of the metastable and unstable state.

The numerical coefficients appearing in Eqns. (\ref{eqn8}), (\ref{eqn15}), (\ref{eqn20}), (\ref{eqn21}), (\ref{eqn22}), (\ref{eqn23}), (\ref{eqn25}), and (\ref{eqn26}) are correct up-to 6 digits after the decimal points. Eventually, all the results we obtained are correct up-to 6 digits after the decimal points. Exact values of the expressions corresponding to these coefficients, however, can be obtained from the Riemann zeta functions and the series summations in Eqns. (\ref{eqn7}), (\ref{eqn14}), (\ref{eqn17}), and (\ref{eqn17b}).

The Hartree-Fock-Bogoliubov approximation, however, works somewhat better than the Hartree-Fock approximation (which though is extensively used \cite{Giorgini2,Biswas2}) near the critical regime \cite{Davis2}. A finite temperature scaling theory for the collapse within the Hartree-Fock-Bogoliubov approximation is kept as an open problem.

The collapse of the trapped BEC was also understood in the language of macroscopic quantum tunnelling \cite{Ueda}. The study of the macroscopic quantum tunnelling for the collapse within the finite temperature scaling theory is also kept as an open problem. The collapse of the harmonically trapped BEC has also been theoretically investigated for Rashba-Dresselhaus spin-orbit coupling \cite{Mardonov}. Such an investigation in the presence of the artificial magnetic field is also kept as an open problem.

Finally, we have studied the criticality of specific heat ($C_v\propto|T-T_c|^{\alpha}$) for $N=N_c$. The study of the critical phenomena with other thermodynamic variables such as the (local) isothermal compressibility at $\vec{r}=0$ for $N=N_c$ is also kept as an open problem. The critical phenomena for the gravitational collapse in the areas of Gravitation and Cosmology is also a subject matter of study for a completely different system \cite{Gundlach}. The study of the finite temperature scaling theory in such a case is also kept as an open problem.

\section*{Data availability statement}

No new data were created or analysed in this study.

\section*{Acknowledgement}

S. Biswas acknowledges the partial financial support of the SERB (now ANRF), DST, Govt. of India under the EMEQ Scheme [No. EEQ/2023/000788]. Useful discussion with Prof. J. K. Bhattacharjee (HRI, Allahabad, India) is gratefully acknowledged. We thank the anonymous reviewer for his/her thorough review and highly appreciate his/her comments and suggestions which have significantly contributed to improving the quality of the article.


\begin{thebibliography}{99}
\bibitem{Num}For this value of $\nu=0.982341$, energy of the system (thermal cloud) at $T=T_c$ is minimized for $a_s/\bar{l}=-0.0066$ and $N=10^8$.
\bibitem{Num2}For this value of $\nu=1.01622$, energy of the system (thermal cloud) at $T=T_c$ is minimized for $a_s/\bar{l}=0.0066$ and $N=10^8$.
\end{thebibliography}
\end{document}